# Analytical study of the plasmonic modes of metal nanoparticle circular array

Kin Hung Fung and C. T. Chan

Department of Physics, The Hong Kong University of Science and Technology,

Hong Kong, China

We analyze the plasmonic modes of a metal nanoparticle circular array. Closed form solutions to the eigenmode problem are given. For each polarization, the plasmonic mode with the highest quality is found to be an antiphase mode. We found that the significant suppression in radiative loss can be understood as the cancellation of the dipolar radiation term in the radiative linewidth. The remaining finite radiative linewidth decreases exponentially as the number of particle increases.

## I. Introduction

Subwavelength phenomena of electromagnetic waves have attracted considerable attention in recent years. Plasmonic materials are capable of supporting subwavelength phenomena near optical frequencies, and as such, they have been a focus of recent research in near-field optics [1,2]. The tremendous growth of interest in plasmonic materials in form of matel nanoparticles (MNPs) are fueled by the significant improvement in fabrication techniques [3,4] as well as many plausible applications such as biosensors [5], surface enhanced Raman scattering (SERS) [6,7], subwavelength waveguides [8,9] and near-field imaging [10].

There have been many numerical studies on MNPs with various shapes and arrangements. In particular, it has been shown that MNP circular arrays may serve as electric and magnetic resonators [11] and can significantly reduce the radiative loss [12]. In general, due to the complexity of plasmonic modes, analytical studies were few even if dipole approximation can be employed to MNPs in many cases. Markel [13] obtained the solutions for 2 and 4 particles. Citrin [12] obtained the solution for circular MNP arrays, and by taking into account all dipolar couplings between particles, he showed that nearest-neighbor tight-binding model cannot be applied because it may give non-causal solutions. However, concrete examples and more thorough analytical work on the complicated solutions are necessary for a clear understanding of the problem. One of the purposes of this paper is to study an accurate analytical solution to the problem using concrete examples. Since the derivation of the closed form solutions are not yet published and there exist typos [14] in some of the solutions in Ref. [12], we will also make a rigorous derivation for the correct solutions. By considering the dependence of the actual resonant frequencies on materials and geometry, we will show how the asymptotic forms of the solutions can help us understand the physical problem. The analytical works and discussions will focus on the plasmonic modes that have the low radiative loss. Such high quality plasmonic modes might be useful for good resonators for recently proposed applications [15--17].

The paper will be organized as follows. In section II A, the geometries and the material parameters will be firstly described. Closed form solutions to the plasmonic eigenmodes will be derived in Sec. III A and B in details, followed by a comparison with some numerical results in Sec. III C. Mode qualities of the high quality states will be analyzed in Sec. III D. Conclusion and some general discussions will be given in Sec. IV.

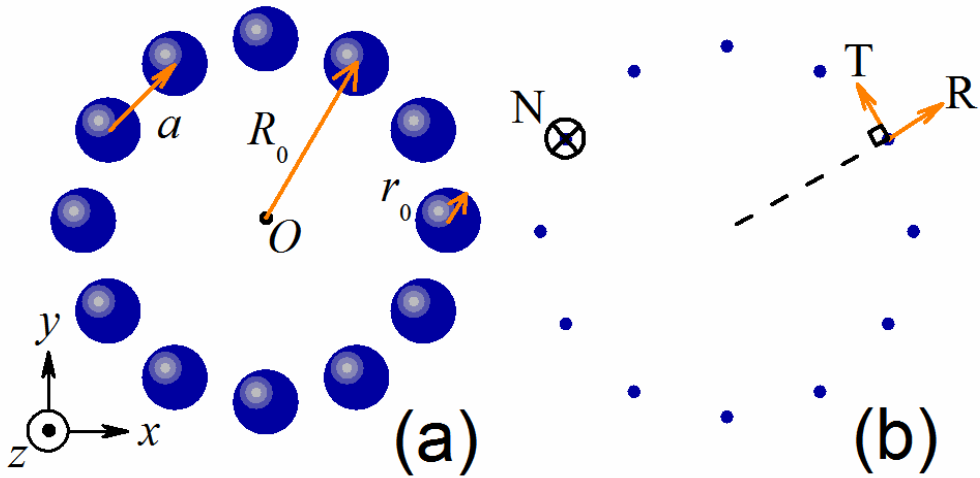

Fig. 1: (Color online) Schematic diagram of the MNP circular array. (a) Geometrical parameters (b) Polarization vectors.

## II. Geometries and material parameters

We consider a circular array of spherical MNPs of particle radius $r_0$ [see Fig. 1(a)]. Taking the center of the circular array as the origin, the position vectors of the particle centers are

$$\mathbf{R}_m = R_0 \cos\frac{2m\pi}{N}\hat{\mathbf{x}} + R_0 \sin\frac{2m\pi}{N}\hat{\mathbf{y}}, \qquad (1)$$

where $m = 0, 1, 2, ..., N-1$ is the particle index, $R_0$ is the radius of the circular array, and $N$ is the number of particles. The particle center-to-center distance is $a = 2R_0 \sin\frac{\pi}{N}$. The dielectric function of the MNPs is taken to be of the Drude's form:

$$\varepsilon(\omega) = 1 - \frac{\omega_p^2}{\omega(\omega + i\gamma)}, \qquad (2)$$

where $\omega$ is the angular frequency, $\omega_p$ is the plasma frequency, and $\gamma$ is the electron scattering rate. The background medium has a dielectric constant denoted by $\varepsilon_h$. We assume that there is no magnetic permeability everywhere (i.e. $\mu = \mu_h = 1$). We will restrict ourselves to the situations that the MNP are not too close together, so that $a \geq 3r_0$, which is a customary assumption for the validity of the dipole approximation.

## III. Eigenmode analysis

To facilitate the understanding of the electromagnetic resonances of MNP circular array, we may solve for the intrinsic normal modes of such system. The analysis is complicated by the fact that there is always radiation energy loss in an finite array of particles. On top of radiation loss, absorption loss is not negligible for plasmonic system. To describe the electromagnetic resonances of such kind of systems, a good approach is the eigen-decomposition (ED) method [13,18,19] (sometimes called spectral theory). Bergman *et. al.* [18] have proposed such a method for the understanding of electromagnetic resonances of a finite number of spherical objects. The theory considers all multipolar responses of a spherical object. Later, a simplified version of method that only considers the dipolar response of small particles was developed by Markel *et. al.* [13]. Such dipolar ED method was applied successfully to systems such as fractal clusters, and periodic MNP arrays. An important advantage of such approach is that it does not require either numerical complex root searching [20--22] or root approximation [12]. Here, we will employ the dipolar ED method to analyze the response of MNP circular arrays. We will show that the method allow us to obtain analytical solutions.

**A. Dynamic dipole eigenvalue equation**

We consider the case where the particles are well separated ($a \geq 3r_0$) such that dipole approximation can already make a good description in optical frequencies. We note that the only approximation that we use in this paper is the dynamic dipole approximation [23,24]. Retardation effects will be included through the dynamic Green's function and material polarizability.

If there is an external time-periodic driving electric field, $\mathbf{E}_m^{ext} e^{-i\omega t}$, acting on the $m$-th particle, the coupled dipole equations can be written as

$$\mathbf{p}_m = \alpha \left[ \mathbf{E}_m^{ext} + \sum_{n \neq m} \overset{\leftrightarrow}{\mathbf{W}}(\mathbf{R}_m - \mathbf{R}_n) \mathbf{p}_n \right], \quad (3)$$

where the dynamic Green's function is

$$W_{uv}(\mathbf{r}) = k_0^3 [A(k_0 r) \delta_{uv} + B(k_0 r) \frac{r_u r_v}{r^2}], \quad (4)$$

$$A(x) = (x^{-1} + ix^{-2} - x^{-3}) e^{ix}, \quad (5)$$

$$B(x) = (-x^{-1} - 3ix^{-2} + 3x^{-3}) e^{ix}, \quad (6)$$

$k_0 = \omega/c$, $c$ is the speed of light in the background medium, and $u, v = 1, 2, 3$ are component indices in Cartesian coordinates. Equation (3) can be written as

$$\sum_{n=1}^{N} \sum_{v=1}^{3} M_{mnuv} p_{nv} = E_{mu}^{ext}, \quad (7)$$

or, in matrix form, $\mathbf{Mp} = \mathbf{E}$, Here, $\mathbf{M}$ can be divided into two parts so that

$$\mathbf{M} = \alpha^{-1} \mathbf{I} - \mathbf{G}, \quad (8)$$

where $\mathbf{I}$ is the identity matrix, and

$$G_{mnuv} = \begin{cases} W_{uv}(\mathbf{r}_m - \mathbf{r}_n), & m \neq n \\ 0, & m = n, \end{cases} \quad (9)$$

which is independent of the material properties. The polarizability, $\alpha$, is the dynamic dipole polarizability [23,24], which is given by [25]

$$\alpha(\omega) = i\frac{3c^3}{2\omega^3}a_1(\omega). \qquad (10)$$

Here,

$$a_1(\omega) = \frac{q\psi_1(qx)\psi_1'(x) - \psi_1(x)\psi_1'(qx)}{q\psi_1(qx)\xi_1'(x) - \xi_1(x)\psi_1'(qx)} \qquad (11)$$

is the "$\ell = 1$" electric term of the Mie's coefficients [26], $\psi_1$ and $\xi_1$ are the Riccati-Bessel functions, $x = \omega a/c$ and $q = \sqrt{\varepsilon(\omega)/\varepsilon_h}$. For analyzing the resonances of cluster of dipoles, the ED method considers the following eigenvalue problem:

$$\mathbf{Mp} = \lambda \mathbf{p}, \qquad (12)$$

where $\lambda$ and $\kappa$ are, respectively, the complex eigenvalues of $\mathbf{M}$ and $\mathbf{G}$, and are related by $\lambda = \alpha^{-1} - \kappa$. The eigenvalues, $\lambda$, and the eigen-polarizability defined by [24]

$$\alpha_{eig} = \frac{1}{\lambda}, \qquad (13)$$

are useful quantities for analyzing the intrinsic normal modes of the system. We will solve such eigenvalue problem for the circular array in the following subsections.

**B. Analytical eigensolutions**

In general, the eigenvalue problem [Eq. (12)] cannot be solved analytically when there are many particles involved ($N \gg 1$) [27]. Also, even if Eq. (12) is solved numerically, it is very difficult to pick a particular eigenmode from a bunch of eigenmodes because the number of eigenmodes is proportional to $N$. However, the problem can be simplified if a system has a high symmetry [24]. The circular array that we considered has discrete rotational symmetry, which allows us to obtain analytical results for the eigenvectors and eigenvalues.

In Eq. (12), the eigenvalue problem is written in a unique set of Cartesian coordinates for each dipole. For this particular problem, it is more convenient to use a different set of coordinates for each dipole. The new vectors after such transformation are

$$p'_{mu} = \sum_{v=1}^{3} \widetilde{\Omega}_{muv} p_{mv} \qquad (14)$$

(or $\mathbf{p}' = \mathbf{\Omega}\mathbf{p}$ in matrix form), where the local transformation for the $m$-th dipole is

$$\tilde{\mathbf{\Omega}}_m = \begin{pmatrix} \cos\dfrac{2m\pi}{N} & \sin\dfrac{2m\pi}{N} & 0 \\ -\sin\dfrac{2m\pi}{N} & \cos\dfrac{2m\pi}{N} & 0 \\ 0 & 0 & 1 \end{pmatrix} \tag{15}$$

and the global transformation for the whole system is

$$\mathbf{\Omega} = \begin{pmatrix} \tilde{\mathbf{\Omega}}_1 & \mathbf{O} & \mathbf{O} & \mathbf{O} & \mathbf{O} \\ \mathbf{O} & \tilde{\mathbf{\Omega}}_2 & \mathbf{O} & \mathbf{O} & \mathbf{O} \\ \mathbf{O} & \mathbf{O} & \tilde{\mathbf{\Omega}}_3 & \mathbf{O} & \mathbf{O} \\ \mathbf{O} & \mathbf{O} & \mathbf{O} & \ddots & \vdots \\ \mathbf{O} & \mathbf{O} & \mathbf{O} & \cdots & \tilde{\mathbf{\Omega}}_N \end{pmatrix}. \tag{16}$$

The transformed eigenvalue problem becomes

$$\mathbf{M}'\mathbf{p}' = \lambda \mathbf{p}', \tag{17}$$

where $\mathbf{M}' = \mathbf{\Omega}\mathbf{M}\mathbf{\Omega}^{-1}$. Due to the discrete rotational symmetry of the system, $\mathbf{M}'$ is invariant under a cyclic index-translation $m \to Mod(m+n, N)$, i.e. $\mathbf{T}(n)\mathbf{M}'\mathbf{T}(n)^{-1} = \mathbf{M}'$, for $n = 1,2,3,...,N$, where

$$\mathbf{T}(n) = \begin{pmatrix} \mathbf{O} & \tilde{\mathbf{I}} & \mathbf{O} & \mathbf{O} & \mathbf{O} \\ \mathbf{O} & \mathbf{O} & \tilde{\mathbf{I}} & \ddots & \vdots \\ \mathbf{O} & \mathbf{O} & \mathbf{O} & \ddots & \mathbf{O} \\ \mathbf{O} & \mathbf{O} & \mathbf{O} & \ddots & \tilde{\mathbf{I}} \\ \tilde{\mathbf{I}} & \mathbf{O} & \mathbf{O} & \cdots & \mathbf{O} \end{pmatrix}^n \tag{18}$$

and $\tilde{\mathbf{I}}$ is a $3\times 3$ identity matrix. In other words, $\mathbf{T}(n)$ and $\mathbf{M}'$ commute. We note that the set, $\{\mathbf{T}(1), \mathbf{T}(2), \mathbf{T}(3),..., \mathbf{T}(N)\}$, form a cyclic group $C_N$. The simultaneous eigenvectors of the group and the corresponding eigenvalues of $\mathbf{T}(n)$ are denoted by $\mathbf{v}^{(j)}$ and $t_n^{(j)}$, respectively, with

$$t_n^{(j)} = e^{i2\pi jn/N} \tag{19}$$

and

$$v_{mu}^{(j)} = c_u^{(j)} e^{i2\pi jm/N}, \tag{20}$$

where $c_1^{(j)}$, $c_2^{(j)}$, and $c_3^{(j)}$ are arbitrary constants and $j=1,2,3...,N$. The eigenvectors $\mathbf{v}^{(j)}$, of $\mathbf{T}(n)$ are also the eigenvectors of $\mathbf{M}'$. Therefore, we can write the eigenvectors, $\mathbf{p}'^{(j,\sigma)}$, of $\mathbf{M}'$ in the form of

$$p'^{(j,\sigma)}_{mu} = c_u^{(j,\sigma)} e^{i2\pi jm/N}, \qquad (21)$$

where $\sigma$ is an index allowing further reduced eigenspaces. After substituting (21) into (17), the eigenvalue problem becomes

$$\sum_{v=1}^{3} \widetilde{M}'^{(j)}_{uv} c_v^{(j,\sigma)} = \lambda^{(j,\sigma)} c_u^{(j,\sigma)}, \qquad (22)$$

where

$$\widetilde{M}'^{(j)}_{uv} = \sum_{m=1}^{N} M'_{Nmuv} e^{i2\pi jm/N}. \qquad (23)$$

If we write Eq. (22) explicitly in matrix form, we will finally get the $3\times 3$ matrix equation,

$$\widetilde{\mathbf{M}}'^{(j)} \mathbf{c}^{(j,\sigma)} = \lambda^{(j,\sigma)} \mathbf{c}^{(j,\sigma)}, \qquad (24)$$

where $\widetilde{\mathbf{M}}'^{(j)} = \alpha^{-1}\mathbf{I} - \widetilde{\mathbf{G}}^{(j)}$,

$$\widetilde{\mathbf{G}}^{(j)} = k_0^3 \sum_{m=1}^{N-1} \left[ A(k_0 D_m) \widetilde{\mathbf{\Omega}}_m^{-1} + B(k_0 D_m)(\widetilde{\mathbf{\Omega}}_m^{-1} + \mathbf{K})/2 \right] e^{i2\pi jm/N}, \qquad (25)$$

$D_m = 2R\sin(m\pi/N)$, and

$$\mathbf{K} = \begin{pmatrix} -1 & 0 & 0 \\ 0 & 1 & 0 \\ 0 & 0 & -1 \end{pmatrix}. \qquad (26)$$

The eigenvalues of $\widetilde{\mathbf{G}}^{(j)}$ are found to be

$$\kappa^{(j,1)} = \frac{k_0^3}{2}\left(\Sigma_{jC} + \sqrt{\Sigma_{jB}^2 - \Sigma_{jS}^2}\right), \qquad (27)$$

$$\kappa^{(j,2)} = \frac{k_0^3}{2}\left(\Sigma_{jC} - \sqrt{\Sigma_{jB}^2 - \Sigma_{jS}^2}\right), \qquad (28)$$

$$\kappa^{(j,3)} = k_0^3 \Sigma_{jA} \qquad (29)$$

with the corresponding eigenvectors

$$\mathbf{c}^{(j,1)} = \left(\Sigma_{jB} - \sqrt{\Sigma_{jB}^2 - \Sigma_{jS}^2},\ -\Sigma_{jS},\ 0\right)^{\mathrm{T}}, \qquad (30)$$

$$\mathbf{c}^{(j,2)} = \left(\Sigma_{jB} + \sqrt{\Sigma_{jB}^2 - \Sigma_{jS}^2},\ -\Sigma_{jS},\ 0\right)^{\mathrm{T}}, \tag{31}$$

$$\mathbf{c}^{(j,3)} = (0,\ 0,\ 1)^{\mathrm{T}}, \tag{32}$$

where

$$\Sigma_{jA} = \sum_{m=1}^{N-1} A(k_0 D_m) e^{i 2\pi j m / N}, \tag{33}$$

$$\Sigma_{jB} = \sum_{m=1}^{N-1} B(k_0 D_m) e^{i 2\pi j m / N}, \tag{34}$$

$$\Sigma_{jC} = \sum_{m=1}^{N-1} [2A(k_0 D_m) + B(k_0 D_m)] \cos\frac{2m\pi}{N} e^{i 2\pi j m / N}, \tag{35}$$

$$\Sigma_{jS} = \sum_{m=1}^{N-1} [2A(k_0 D_m) + B(k_0 D_m)] \sin\frac{2m\pi}{N} e^{i 2\pi j m / N}. \tag{36}$$

Finally, the eigenvalues in Eq. (15) are given by

$$\lambda^{(j,\sigma)} = \alpha^{-1} - \kappa^{(j,\sigma)}, \tag{37}$$

with the corresponding eigenvectors, $\mathbf{p}'^{(j,\sigma)}$, given by Eq. (21). We note that $\Sigma_{jS}$ can be zero when $N$ is even and $j = N/2, N$. In case of $\Sigma_{jS} = 0$, the first and second eigenvectors should be replaced by $\mathbf{c}^{(j,1)} = (0,1,0)^{\mathrm{T}}$ and $\mathbf{c}^{(j,2)} = (1,0,0)^{\mathrm{T}}$. The corresponding eigenvalues are $\kappa^{(j,1)} = k_0^3(\Sigma_{jC} + \Sigma_{jB})/2$ and $\kappa^{(j,2)} = k_0^3(\Sigma_{jC} - \Sigma_{jB})/2$. From Eqs. (30), (31), and (32), we can classify the eigenmodes into parallel-to-plane (PP) modes ($\sigma = 1,2$) and normal-to-plane (NP) modes ($\sigma = 3$). For each $\sigma$, there is one non-degenerate in-phase mode ($j = N$) for any $N$, and one additional non-degenerate anti-phase mode ($j = N/2$) for even $N$. Other modes are two-fold degenerate. The NP modes are polarized perpendicular to the $xy$-plane while the PP modes are polarized elliptically in the $xy$-plane, except for $j = N/2$ and $j = N$ that correspond to linearly polarization in the $xy$-plane. The "$\sigma = 1$, $j = N$" mode is the in-phase magnetic mode which is the focus of Ref. [11] and the "$\sigma = 3$, $j = N$" mode is the in-phase pure electric radiative mode.

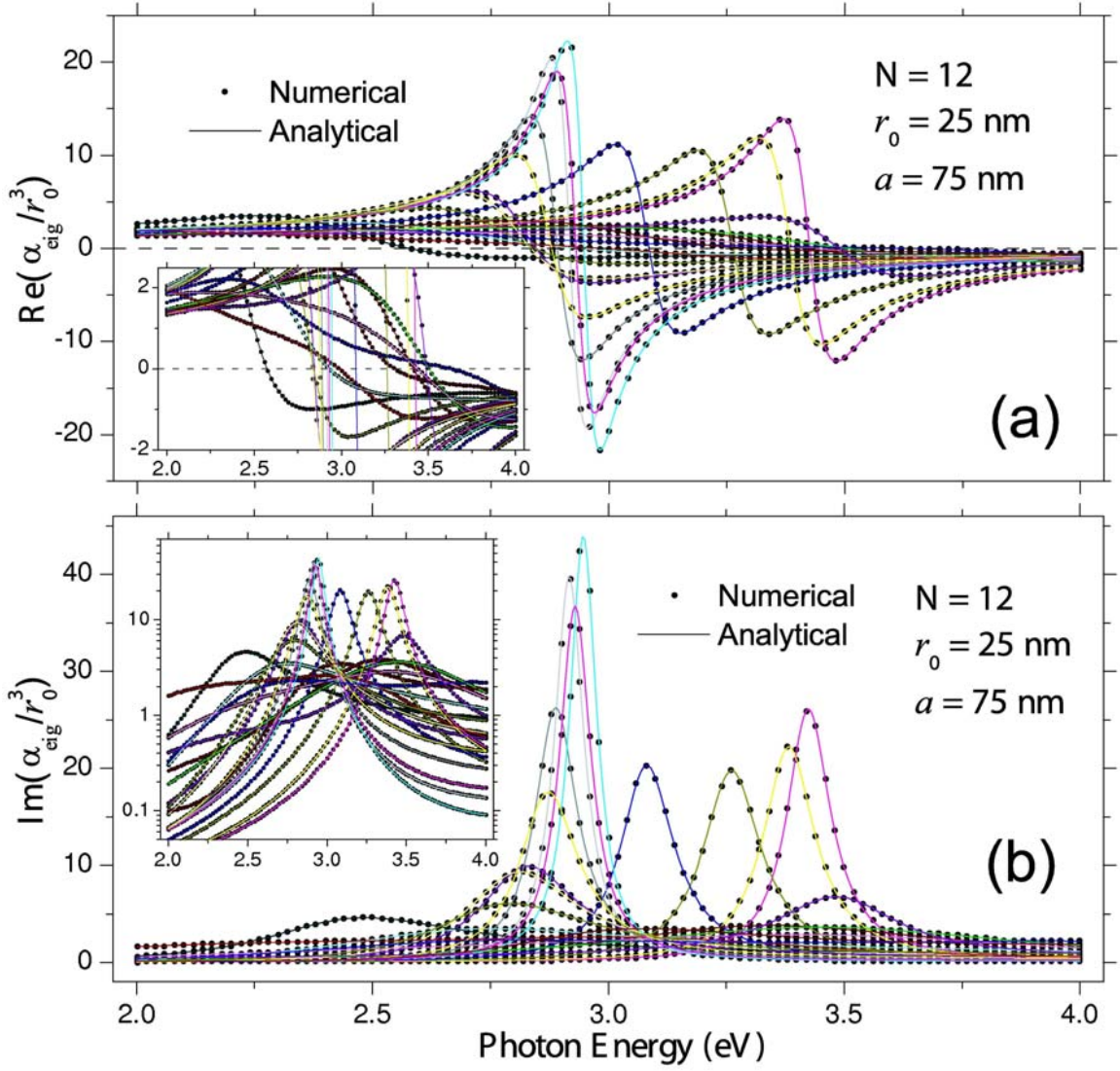

Fig. 2: (Color online) Eigen-polarizabilities versus photon energy. (a) Real part. (b) Imaginary part. Insets show the enlarged parts of the same figure. Parameters are $N=12$, $r_0 = 25$ nm, $a = 75$ nm, $\omega_p = 9.2$ eV, $\gamma = 0.1$ eV, and $\varepsilon_h = 2.5921$.

**C. Polarizabilities of the modes**

In Sec. III B, we have derived the closed form solutions to the eigenvalue problem, Eq. (12). Here, we analyze the eigenmodes using the "eigen-polarizabilities", which is defined in Eq. (13), which has the physical meaning of the collective response function of the whole system for an external electric field pattern that is proportional to the corresponding eigenvector. For a concrete example, we take $N=12$, $r_0 = 25$ nm, $a = 75$ nm, $\omega_p = 9.2$ eV, $\gamma = 0.1$ eV, and $\varepsilon_h = 2.5921$. As a spectral function, the real part and imaginary part of $\alpha_{eig}$, calculated using the analytical formulae Eq. (27), (28), and (29), are shown in Fig. 2. In the same figure, we also include the numerical data calculated by numerical diagonalizations of **M**. The insets of the figure show the enlarged part of the same plot for clarity of the dense region. The analytical and numerical results agree very well. From Fig. 2, we see that there are only 21 curves, and the number of curves is smaller than the dimension of the matrix, $3N (= 36)$. This is consistent with the analysis in Sec. III B that there are 5 two-fold degenerate eigenspaces and 2 non-degenerate eigenvector for each $\sigma$. Although degeneracy exists, the number of curves (~ the complexity) is almost proportional to the number of particles. For large number of particles, it is very difficult to trace the curves from the mixed data. For example, we would not be able to connect the points in Fig. 2 if we do not know the analytic results. In this case, the analytical formulae would help a lot.

We can see a typical Lorenzian resonant feature for each of the eigen-polarizabilities. The real part, $\text{Re}(\alpha_{eig})$, in general exhibits a peak followed by a dip and changes from positive to negative while the imaginary part, $\text{Im}(\alpha_{eig})$, is positive definite. The peaks shown in Fig. 2(b) are the resonant (normal mode) frequencies. The corresponding widths of the peaks, as discussed in more details in Sec. III D, are inversely proportional to the quality of the system. At a first glance, we see a high density of high quality states from $2.9\,\text{eV}$ to $3.5\,\text{eV}$, especially near $2.9\,\text{eV}$. These correspond to the higher order modes of the circular array. The two highest peaks near 2.9 eV correspond to one PP mode and one NP mode with $j = N/2$ while the highest peak near 3.4 eV correspond to another PP mode with $j = N/2$. We will analyze some of those high quality states in Sec. III D.

**D. Mode qualities**

The imaginary part of the eigen-polarizability, $\text{Im}(\alpha_{eig})$, is proportional to the energy extinction. A particular peak frequency of the spectral function, $\text{Im}(\alpha_{eig})$ indicates a resonant frequency, $\omega_0$, and the corresponding peak width, $\delta$, equals to the inverse of relaxation time, $1/\tau$. The quality factor, $Q$, of a particular mode is then given by $\omega_0 / \delta$. To find $Q$ analytically, it is more convenient to find the form of $\text{Im}(\lambda)$, which is related to $\text{Im}(\alpha_{eig})$ by

$$\text{Im}(\alpha_{eig}) = -\frac{\text{Im}(\lambda)}{\text{Re}(\lambda)^2 + \text{Im}(\lambda)^2}. \tag{38}$$

Near the resonant frequency, we can approximate $\text{Re}(\lambda)$ by

$$\text{Re}[\lambda(\omega)] \approx (\omega - \omega_0)\chi(\omega_0), \tag{39}$$

where $\chi(\omega) = \dfrac{d \, \text{Re}[\lambda(\omega)]}{d\omega}$ since $\text{Re}[\lambda(\omega)]$ will cross zero at resonance [28]. With (38) and (39), we thus have

$$\delta \approx -\frac{\text{Im}[\lambda(\omega_0)]}{|\chi(\omega_0)|}. \tag{40}$$

The function, $\chi(\omega)$, depends very much on the choice of material. In this subsection, we will not analyze the effect of energy absorption by a particular material and will focus mainly on the suppression of radiation loss. For simplicity, we do not include absorption in the following derivations. Since $\chi(\omega_0)$ is approximately the same for all modes, $\text{Im}[\lambda(\omega_0)]$ is usually used to analyze the linewidth [13]. It can be divided into two parts:

$$\text{Im}(\lambda) = -\frac{2}{3}k_0^3 - \text{Im}(\kappa), \tag{41}$$

Here, we have used the fact that $\text{Im}(1/\alpha) = -2k_0^3/3$ for non absorbing particles. For large $N$, we may take $N$ to be even, the anti-phase modes ($j = N/2$), where each pair of nearest dipoles have opposite dipole moments, will have the smallest linewidth for each $\sigma$. We thus focus on the "$j = N/2$" modes. The dipole moments of these modes are linearly polarized along the tangential (T), radial (R), and normal (N) direction for $\sigma = 1$, 2, and 3 respectively [see Fig. 1(b)]. The corresponding eigenvalues, given by Eq. (37), can be written as

$$\kappa_T = k_0^3 (\Sigma_{long} + \Sigma_{cross}), \tag{42}$$

$$\kappa_R = k_0^3 (\Sigma_{trans} + \Sigma_{cross}), \tag{43}$$

$$\kappa_N = k_0^3 \Sigma_{trans}, \tag{44}$$

where $\Sigma_{trans}$, $\Sigma_{long}$, and $\Sigma_{cross}$ comes from transverse, longitudinal, and crossed interactions. Using Laurent series expansion, we get

$$\Sigma_{trans} = \sum_{n=-3}^{\infty} \frac{(n+2)^2}{(n+3)!} \frac{i^{n+1} S_{N,n}}{\sin^n \frac{\pi}{N}} (k_0 a)^n, \tag{45}$$

$$\Sigma_{long} = \sum_{n=-3}^{\infty} \frac{2(n+2)}{(n+3)!} \frac{i^{n+1} S_{N,n}}{\sin^n \frac{\pi}{N}} (k_0 a)^n, \tag{46}$$

$$\Sigma_{cross} = \sum_{n=-3}^{\infty} \frac{(n+2)(n+4)}{2(n+3)!} \frac{i^{n+3} S_{N,n+2}}{\sin^n \frac{\pi}{N}} (k_0 a)^n, \tag{47}$$

where

$$S_{N,n} = \sum_{m=1}^{N-1} (-1)^m \sin^n \frac{m\pi}{N}. \tag{48}$$

If we take the imaginary part of the eigenvalues, we only have terms with $n = 0, 2, 4, \ldots$. Further more, for these even terms, we have

$$S_{N,n} = \begin{cases} -1 & \text{for } n = 0 \\ 0 & \text{for } n = 2, 4, 6, \ldots, N-2 \\ N(-1)^{N/2} 2^{1-n} C_{\frac{n-N}{2}}^n & \text{for } n = N, N+2, N+4, \ldots, 3N-2 \end{cases}. \tag{49}$$

Finally, we have

$$\frac{\text{Im}[\kappa_T]}{k_0^3} = -\frac{2}{3} + \frac{N^2(N+2)}{2^{N-2}(N+1)! \sin^{N-2} \frac{\pi}{N}} (k_0 a)^{N-2} + \ldots, \tag{50}$$

$$\frac{\text{Im}[\kappa_R]}{k_0^3} = -\frac{2}{3} + \frac{N^2(N+2)}{2^{N-2}(N+1)! \sin^{N-2} \frac{\pi}{N}} (k_0 a)^{N-2} + \ldots, \tag{51}$$

$$\frac{\text{Im}[\kappa_N]}{k_0^3} = -\frac{2}{3} + \frac{N(N+2)^2}{2^{N-1}(N+3)! \sin^N \frac{\pi}{N}} (k_0 a)^N + \ldots. \tag{52}$$

{We note that Eq. (50) and (51) are the same only up to the term associated with $(k_0 a)^{N-2}$.} These formulae are generalized versions (for $N$ particles) of the expression derived by Markel for two and four particles [13].

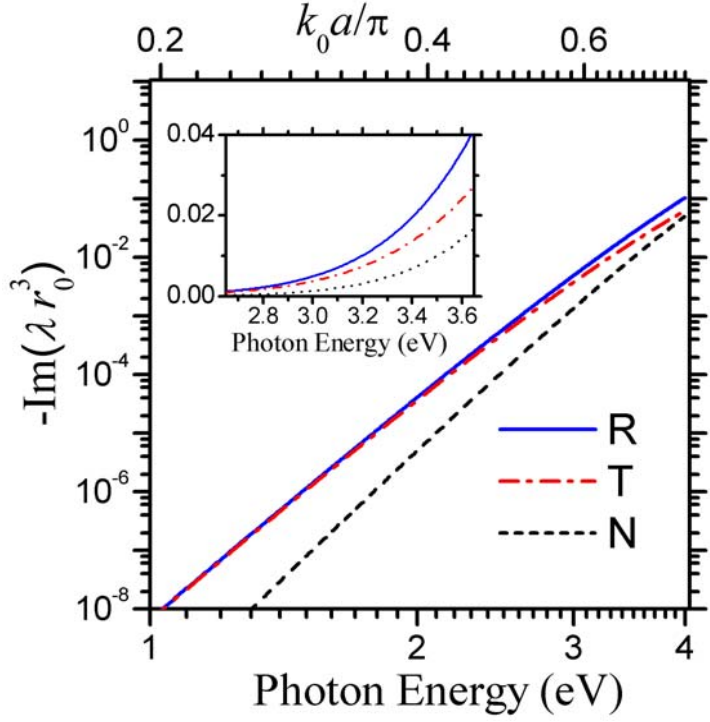

Fig. 3: (Color online) Linewidths of the highest-quality plasmonic modes for each polarization versus resonant frequency. Radial mode (R). Tangential mode (T). Normal-to-plane mode (N). Inset shows an enlarged part in linear scale. $\gamma = 0$.

We see that all of the above expressions have the first term, $-2/3$, that can exactly cancel the single dipole radiation term in Eq. (41), i.e. the collective dipolar radiation is suppressed. The remaining higher order terms contribute to the linewidth. For a fixed $N \gg 1$, as long as $k_0 a$ is small enough such that the first non-zero higher order term dominates, the linewidths are $\delta \sim (k_0 a/\pi)^{N+1}$ for both tangential and radial modes and $\delta \sim (k_0 a/\pi)^{N+3}$ for normal-to-plane mode. Compared with the linewidth of dipole resonance, $\delta \sim (k_0 a)^3$, the linewidths of these anti-phase modes are significantly reduced when $N$ is large. The above approximate forms of the linewidths are accurate when $k_0 a \ll \pi$. For an example, we take the same parameters as in Sec. C except $\gamma = 0$. Fig. 3 shows a plot of the exact analytical value of $-\mathrm{Im}(\lambda r_0^3)$ [given by Eq. (37)] against resonant frequency. At low frequency ($k_0 a < 0.4\pi$), the graph (in log-log scale) shows straight lines with slopes that are consistent with the approximated power factor $N+1$ for tangential and radial modes, and $N+3$ for normal-to-plane mode. The overlapping between the tangential and radial modes also agree with the expression in Eqs. (50) and (51). The lines bend slightly downwards at higher frequencies ($k_0 a > 0.4\pi$), indicating that the linewidths are smaller than the approximated power laws. In this range, the tangential and radial modes also start to separate from each other. In general, the linewidth of a resonant state decreases rapidly as the resonant frequency decreases. The actual resonant frequencies fall in the range of $0.4\pi < k_0 a < 0.7\pi$ for our material parameters. By inspecting the interaction of surface charges [29] on the MNPs, we know that the normal-to-plane and radial modes have similar resonant frequencies that are lower than the single sphere resonant frequency while the tangential mode is higher. Therefore, this suggests the linewidth of tangential mode is much larger than that of the radial mode although their linewidths have the same form [$\delta \sim (k_0 a / \pi)^{N+1}$]. Furthermore, the linewidth of normal-to-plane mode is smaller than that of the tangential mode by a factor of $(k_0 a / \pi)^2$ approximately. Fig. 4 shows that exact analytical values of Im($\alpha_{eig}$) for these three modes. The figure show sharp peaks that agree with our estimation in terms of both peak frequency and peak width. Here, we conclude that the best quality mode is the anti-phase NP mode with a linewidth that takes the form of $\delta \sim (k_0 a / \pi)^{N+3}$.

Although $N$ is assumed to be even in the above analysis, the general properties of the linewidths are almost the same for odd $N$. However, the linewidth of the highest quality state for odd $N$ is $\delta \sim (k_0 a / \pi)^{N+2}$ instead. Another difference is that the highest quality state(s) for even $N$, i.e. the "$j = N/2$" state, is non-degenerate while that for odd $N$, i.e. the "$j = (N \pm 1)/2$" states, are two-fold degenerate. Fig. 5 shows the linewidths [given by Eq. (37)] of the highest quality modes against resonant frequency for different number of particles. We see that the linewidths increase with the resonant frequency in the form very close to our predicted power laws. For a given even number $N_0$, the linewidths of the "$N = N_0$" case and the "$N = N_0 - 1$" case nearly overlap. In addition, it is shown in Fig. 6 that the linewidths decay exponentially as $N$ increases (separately for even and odd numbers).

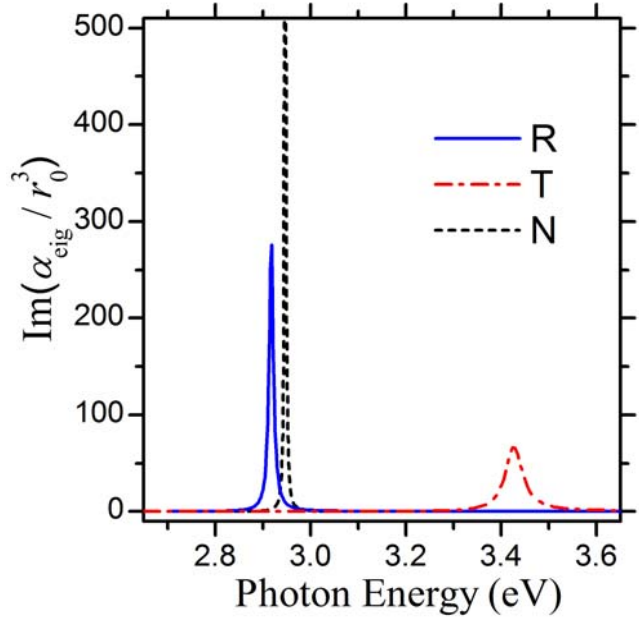

Fig. 4: (Color online) $\text{Im}(\alpha_{eig})$ of the highest-quality plasmonic modes for each polarization versus photon energy. Radial mode (R). Tangential mode (T). Normal-to-plane mode (N). $\gamma = 0$.

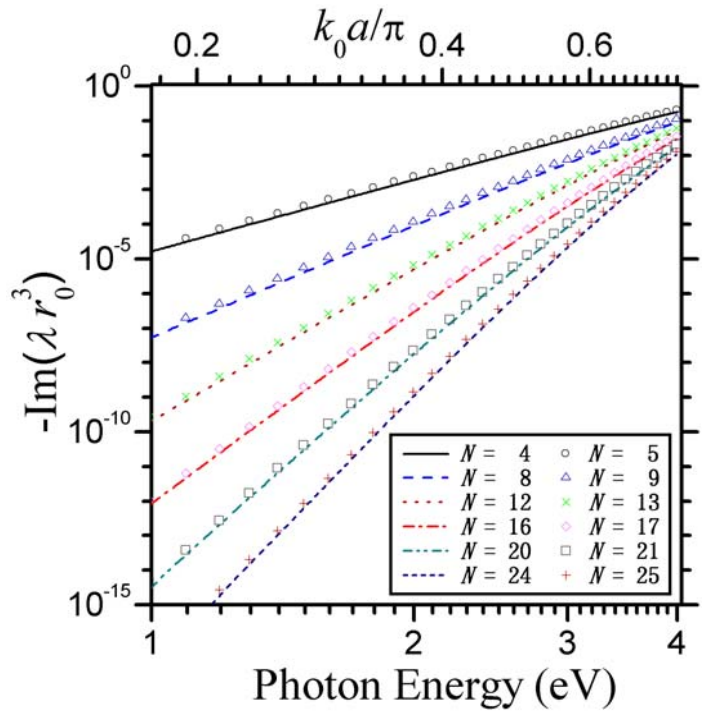

Fig. 5: (Color online) Linewidths of the normal-to-plane mode versus resonant frequency. $\gamma = 0$.

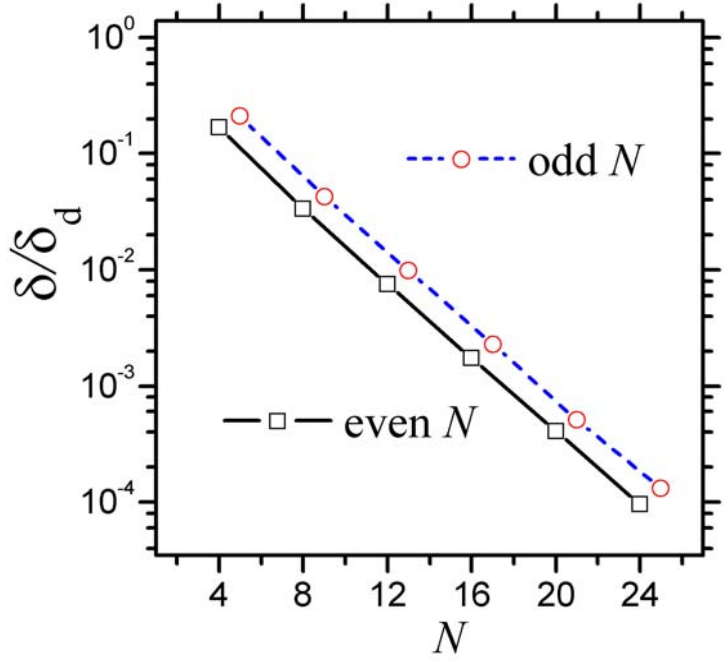

Fig. 6: (Color online) Linewidths of the normal-to-plane mode versus number of particles. $\gamma = 0$.

## IV. Discussion and conclusion

In summary, we have rigorously derived the analytical eigen-solutions to the plasmonic modes of MNP circular array within the dynamic dipole approximation. Significant suppression of radiation loss was found in some of the eigenmodes. Those high-quality plasmonic modes were understood by using the analytical solutions. We found that the anti-phase mode could almost cancel the radiative linewidth of single dipole. The normal-to-plane anti-phase mode was found to be the highest quality mode among all modes. If the resonant frequency is low ($k_0 a < \pi$), the remaining radiative linewidth was found to be dominated by a term proportional to $(k_0 a / \pi)^{N+3}$ for even number of particles, $N$. Our solutions could be useful in understanding the complicated response properties of MNP circular array and wave propagation in MNP plasmonic waveguide with bending.

Our results shown in this paper are based on local theory, but one could easily rederive the results using nonlocal theory, provided that a valid non-local dielectric function is given.

## ACKNOWLEDGMENTS

This work was supported by the Central Allocation Grant from the Hong Kong RGC through HKUST 3/06C. Computation resources were supported by the Shun Hing Education and Charity Fund. We also thank Dr. Jianwen Dong and Dr. Dezhuan Han for useful discussions.